\documentclass[12pt,preprint]{aastex}

\shorttitle{A 3~$\mu$ study of AFGL~2688}
\shortauthors{Goto et al.}

\begin{document}

\title{Imaging and Spatially Resolved Spectroscopy of AFGL 2688 in
the Thermal Infrared Region \footnote{Based on data collected at Subaru
Telescope, which is operated by the National Astronomical Observatory
of Japan.}}

\author{Miwa Goto\footnote{Visiting astronomer at the Institute for
Astronomy, University of Hawaii}, Naoto Kobayashi, and Hiroshi Terada}
\affil{Subaru Telescope, 650 North A`ohoku Place, Hilo, HI 96720}
\email{mgoto@duke.ifa.hawaii.edu} \and

\author{A. T. Tokunaga}
\affil{Institute for Astronomy, University of Hawaii, 
    Woodlawn Dr., Honolulu, HI 96822}

\begin{abstract}
We present ground-based high-resolution ($\sim$0\farcs 3) imaging of
AFGL~2688 at $L'$ (3.8~\micron) and $M'$ (4.7~\micron).  A wealth of
structure in the central region is revealed due to less extinction in
the thermal infrared. A clear border in the southern lobe at $L'$
corresponds to the edge of the heavily obscured region in visible,
indicating there is a dense material surrounding the central
region. The images also show a narrow dark lane oriented to
140$^\circ$ east of north with the normal at 50$^\circ$. The normal
position angle is inconsistent with the optical polar axis (PA =
15$^\circ$), but is aligned to the high-velocity CO components found
in the radio wavelength observations. The central star remains
invisible at $L'$ and $M'$. Several clumpy regions in the north lobe
dominate in $L'$ and $M'$ luminosity. In particular a pointlike source
(peak A) at 0\farcs 5 northeast of the center of the nebula exhibits
the highest surface brightness with a very red spectral energy
distribution (SED). Based on the almost identical SED as adjacent
regions, we suggest that the pointlike source is not self-luminous, as
was proposed, but is a dense dusty blob reflecting thermal emission
from the central star.

We also present spatially resolved slit spectroscopy of the bright
dusty blobs. An emission feature at 3.4~\micron~as well as at
3.3~\micron~ is detected everywhere within our field of view. There is
no spatial variation in the infrared emission feature (IEF) throughout
the observed area (0\farcs 2--1\farcs 5, or 240--1800~AU from the
central source).  The constant flux ratio of the emission feature
relative to the continuum is consistent with the view that the blobs
are mostly reflecting the light from the central star in the
3~\micron~region.

\end{abstract}


\section{Introduction} 

Low- to intermediate-mass stars (2--5~M$_\sun$) experience rapid
transition from the asymptotic giant branch (AGB) to planetary nebula
(PN) at the end of their life.  An evolved star has high-mass loss
when ascending the AGB, but at the tip of the AGB the star loses
almost all the hydrogen envelope and cannot sustain high-mass loss.
It then begins to evolve toward higher temperatures in H-R diagram
while keeping almost constant luminosity as a proto-planetary nebula
(PPN).  The central star of the PPN eventually becomes hot enough
($>$30,000~K) to ionize the surrounding material ejected during the
preceding mass-loss phase, and the PN is reached.  The transition
phase from PPN to PN is as short as about 5000~yr \citep{sch83}.  PPN
share several observational properties in common \citep{kwo93,kwo00},
including a significant infrared excess at far-infrared wavelengths
due to the warm circumstellar dust ($T_{\rm dust}$ = 150--250~K), a
low-velocity molecular outflow (10--20~km s$^{-1}$), a central star
that is typically an F- to G- type supergiant, and extended or bipolar
nebulosity in the visible.

AFGL~2688 is a prototypical carbon-rich PPN, one of the most
extensively studied so far since the discovery of reflection
nebulosity \citep{ney75}.  AFGL~2688 is located 1.0--1.5~kpc away from
us \citep{coh77} (hereafter we assume the distance to AFGL~2688 to be
1.2~kpc). From the visible spectroscopy at the reflection nebulosity,
\citet{cra75} found that AFGL~2688 harbors an F5Ia star in the center
of the nebula. It may have left the AGB only 200~years ago
\citep{ski97}.

In spite of many studies of AFGL~2688, the fundamental structure of
the nebula is still actively debated. The old standard model requires
equatorial density enhancement lying perpendicular to the polar axis
to shape the optical bipolar nebula \citep{yus84,lat93,lop00}. The
polar axis is defined by the optical bipolar nebula and lies at a
position angle of about 15$^\circ$.  However, recent millimeter and
centimeter radio observations have revealed features hard to reconcile
with the standard model. These structures including a high velocity
$^{13}{\rm CO}$ flow \citep{yam95}, a compact expanding shell
\citep{cox00}, and elongated emission \citep{jur00}, and all have a PA
$\approx$ 55$^\circ$. No present model can explain this apparent dual
symmetry in AFGL~2688. Also surprising and unexplained is the
quadrupole shape of the H$_2$ emission \citep{lat93,sah98a}.

The goal of our observations is to better understand these apparently
contradictory morphological features. We infer that the features at PA
$\approx$ 55$^\circ$ are young because they have relatively high
velocity or are compact. To investigate the origin of these features,
we must observe close to the central star. However, the inner
5\arcsec~of the nebula suffers from heavy obscuration, and is
virtually invisible at wavelengths shorter than 2~\micron. Imaging in
the thermal infrared region with a 8 m class telescope is essential to
access the innermost region of AFGL~2688, as well as to provide high
spatial resolution comparable with the visible image.

This paper is organized as follows: In \S 2 we describe the
observations and reduction of the data. In \S 3 we summarize current
understanding of AFGL~2688 and discuss several new and recently found
morphological features.

\section{Observations and Data Reduction}
\subsection{$L'$ and $M'$ Imaging}

The imaging observations were made on UT 2001 July 14 at the Subaru
Telescope with the Infrared Camera and Spectrograph (IRCS)
\citep{tok97e, kob00}. IRCS is a cryogenic camera and spectrograph
equipped with two 1024~$\times$ 1024 InSb arrays. We used special
$L'$- and $M'$-band filters optimized for the atmospheric transmission
curve at Mauna Kea \citep{tok01}.  Images of $L'$ were recorded by
58~mas~pixel$^{-1}$ camera (60\arcsec $\times$ 60\arcsec~field of
view) at five different locations with the array offset by
30\arcsec~to each time so that median sky frames could be made. The
$M'$ images were recorded in the same manner, but with the 22~mas
pixel$^{-1}$ camera (23\arcsec $\times$ 23\arcsec~field of view) and
smaller nodding offset (10\arcsec). The total on-source integration
times were 270 and 360 s in $L'$- and $M'$, respectively.  Flux
calibration was facilitated by observing a photometric reference star,
HD~201941, selected from \citet{eli82}. The seeing was good ($\sim
$0\farcs 3 in FWHM in $L'$), and stable throughout the observations.
The actual spatial resolution at $M'$ is somewhat inferior to the
seeing FWHM because the high-resolution camera of IRCS was internally
slightly out of focus.

The standard image reduction procedure was performed with
IRAF.\footnote{IRAF is distributed by the National Optical Astronomy
Observatories, which are operated by the Association of Universities
for Research in Astronomy, Inc., under cooperative agreement with the
National Science Foundation.} After sky subtraction, flat-fielding by
dividing by a sky flat, and bad pixel correction, image frames were
registered and averaged. We show the final images in Figures
\ref{imglp1} and \ref{imgmp1} in linear and logarithmic scales. Image
convolution was applied with a 2D Gaussian filter of 1.4~pixels FWHM
to show the faint nebulosity.

The $L'$ and $M'$ images look quite similar to each other. In addition
to the spindle features commonly observed at the shorter wavelengths,
they reveal great details in the innermost region for the first
time. The upper part of south lobe close to the center is very faint
in the 2.15~\micron~image taken by {\it HST/NICMOS} \citep{sah98a},
but it is the brightest region in the southern lobe at the thermal
infrared wavelength. The north lobe appears by far brighter near the
center and harbors several bright blobs. A clear nebulosity enclosing
the blobs extends farther to the east of the north lobe somewhat
independently of the spindle shape. The peak surface brightness is
measured at 6.2~mag~arcsec$^{-2}$ and 4.4~mag~arcsec$^{-2}$ at the
southernmost bright blob at $L'$ and $M'$, respectively. The total
magnitudes are $L'$ = 5.5~mag and $M'$ = 3.5~mag.  The central star is
invisible in either wavelength, but a narrow dark lane lying
140$^\circ$ east of north in the center of the nebula was clearly
detected in our image. The faint nebulosity extending about
6\arcsec~east of the north lobe roughly corresponds to the CO
molecular outflow observed at 1.3~mm \citep{cox00}, suggesting shocked
H$_{\rm 2}$ emission falls on the $L'$ filter coverage.

We used {\it HST/NICMOS} archival data of AFGL~2688 obtained with
H$_2$ continuum (2.15~\micron) and $H$-band (1.65~\micron) filters and
our $L'$ image to create a three-color composite image. The three
images were registered at the brightest blob in the northern
lobe. This could be inaccurate because the precise position of the
surface brightness peak can be intrinsically different in each color,
but we did not find any significant offset in the spindle pattern in
the registered three images. The resultant composite image is shown in
Figure \ref{a2hkl1} in different stretches.

\subsection{3~\micron~Spectroscopy}

The spectroscopic observation was made on UT 2000 September 24 with
the same instrument and telescope using the medium-resolution grism in
the IRCS camera section. The slit was put along north to south at
three different locations on the several bright blobs in the northern
lobe. The slit positions are illustrated in Figure \ref{slit2}. We
used the 0\farcs 3 slit to match the excellent seeing of the night
($\sim$0\farcs25 in FWHM) to obtain low-resolution (R = 600--800)
spectra from 2.84 to 4.18~\micron. The sky background was subtracted
by nodding the telescope by 3\arcsec~along the slit length.  Special
attention was paid to stay on one position for no longer than 60 s to
ensure good sky subtraction. The total on-source integration time was
140 s at each slit position. We observed an F5V star (HR~1687, V =
5.0) as a spectroscopic standard at nearly the same airmass to cancel
out the telluric absorption lines. The difference in the airmass
between the object and the standard star was kept no larger than 0.1
so that the telluric transmission lines canceled well. Spectroscopic
flat frames were obtained using a halogen lamp at the end of the
night.

After we performed sky subtraction, flat-fielding, and correction of
bad pixels in 2D spectrogram images, one-dimensional spectra were
obtained using the aperture extraction package of IRAF. The width of
the aperture was 0\farcs 233 (4 pixels), about the same size as the
seeing.  We defined the nine apertures within the slit offset by
0\farcs 233 from each other covering $\pm$1\arcsec~from the center of
the slit (Fig. \ref{slit2}).

Wavelength calibration was performed using over 60 telluric absorption
lines. First, we calculated the model telluric transmission curve with
matched spectral resolution using the ATRAN atmosphere modeling
software \citep{lor92}. Then the local minimums of the model curve
were picked up to make a list of telluric lines. The result of fitting
the observed atmospheric absorption lines with a calculated line list
shows differences no more than $\pm$0.001~\micron~in peak-to-peak all
across the spectra. The intrinsic stellar lines of the spectral
standard (mostly atomic hydrogen) were hard to remove because of the
blending with the telluric absorption lines. Only Br$\alpha$
(4.05~\micron) and Pf$\gamma$ (3.74~\micron) were fit with a
Lorentzian profile and subtracted before dividing. The one-dimensional
spectra were registered to the standard star spectrum by subpixel
shifting. The small discrepancy in the airmass between the standard
and the object was corrected by rescaling the spectra according to
Beer's law. Since the 0\farcs 3 slit width corresponds to 5 pixels,
the spectra shown in Figure \ref{spec1} were binned by 2 pixels. The
spectrum of the standard star was assumed to be a Planck function of
$T_{\rm eff} = $ 6530~K, and the $L'$ magnitude of the standard star
was estimated using the color of an F5V star ($V-L$ = 1.35) in
\citet{tok00}.

\section{Discussion}
\subsection{Current Understanding of the Nebula's Structure}

A summary of our present understanding of AFGL~2688 is sketched in
Figure \ref{sketch1}. We focused our attention on recent observational
studies of the morphology and kinematics of the nebula. A detailed
discussion of models, such as the one by \citet{ski97}, are beyond the
scope of this paper.

In the past decades, this nebula has been the one of the primary
targets for new observing facilities and techniques. The central star
has not been detected to date at any wavelength (but see Cox et
al. 2000, who found a millimeter continuum peak). However, the
location of the illumination source in the nebula is determined
reasonably well by the center of the ``searchlight beams''
\citep{sah98b} and the symmetric center of the polarization map in the
reflection nebula \citep{wei00}. In addition,the polarization center
roughly coincides with the intersection of lines connecting blue- and
redshifted components of multiple CO outflows
\citep{cox00}. Low-velocity (10--20~km~s$^{-1}$) winds associated with
the mass loss have been detected in the study of molecular line
kinematics at radio wavelengths \citep{you92,yam95}. The visible
nebulosity is thought to be a remnant of mass loss when the star was
still in the AGB phase.

In the context of the current standard model, the horns extending out
of the lobes have been interpreted as limb brightening at the inner
wall of the conical opening. However, \citet{sah98b} found the edges
of these searchlight beams are too sharp, and the contrast between
searchlights and regions between the two searchlight beams are too
high, to be reproducible by limb brightening. They argue that the
sharp edge is not a product of a material distribution. They propose
an obscuring ``dust cocoon'' with a nonuniform annular opening through
which the central source is illuminating the entire nebula to produce
the observed intensity distribution of the searchlight beams.

The concentric arcs around the nebula have been taken to be evidence
of material ejected in the periodic mass loss caused by either
unstable pulsation of the latest AGB phase or recent interaction with
a close binary. With faint and detailed structures revealed by {\it
HST/WFPC2}, \citet{sah98b} found the arcs are almost partial circles
with no additional elongation. They argue that the perfect circularity
is not consistent with latitudinal density variations because the arcs
should proceed faster in the lower-density medium to form more
elongated curves along polar axis, which is not apparent in their
images. The absence of latitudinal density variation in Sahai's model
is not consistent with the current standard model.

In the $H$ and $K$ the nebula lobes appears as two spindles capped
with shocked H$_2$ emission blobs \citep{lat93}.  The clear and bright
spindle edge indicates bubbles invading the low-velocity wind sweeping
up the circumstellar material. The axisymmetrical bubbles are probably
infrared counterparts of the moderate-velocity wind
($\sim$40~km~s$^{-1}$) detected in the radio wavelength
\citep{you92}. They argued against radiation pressure acceleration to
drive such a fast outflow by the dragging of dust grains.


The blue and red components of the radio outflow found in $^{13}$CO
observations are found along PA = 60$^\circ$ \citep{yam95}, rather
than PA = 15$^\circ$ in the optical.  \citet{jur00} have discovered
the thermal emission tail of extremely large dust grains ($a \sim$
0.5~cm) at centimeter wavelengths. The emitting region is elliptical
in shape, with the major axis along PA = 53$^\circ$. An expanding
shell-like structure of 2\arcsec~diameter was found by \citet{cox00}
in CO. The shell is slightly elongated to PA = 54$^\circ$. One of the
puzzling features of the nebula is that the axis of symmetry of the
optical and infrared lobes is at PA = 15$^\circ$, but the axis of
symmetry at radio wavelengths is at PA = 53$^\circ$--60$^\circ$.

\subsection{Sharp Edge in the Southern Lobe}

What is recognized first in our three-color composite image
(Fig. \ref{a2hkl1}) is a clear border in the southern lobe with a
striking color change in the infrared. The crosscut at the south edge
along the optical bipolar axis is shown in Figure \ref{sedge1}.  The
$H$ and 2.15~\micron~flux drops sharply at 1\arcsec~south from the
center. The location of the border roughly corresponds to the edge of
the dust cocoon proposed by \citet{sah98a}.
The red color ($K - L'>$ 3) suggests heavy reddening in the
intervening dust cocoon. The sharpness of the limb implies an
expanding cocoon into the circumstellar envelope pushing away the
surrounding material. To allow bulletlike ejection of the bipolar
bubbles, the cocoon would have to have formed after the lobes.

We noticed the spatial variation of the flux in Figure \ref{sedge1} in
$L'$ and $M'$ is almost the same, and the similarity also holds for
between $H$ and 2.15~\micron, but it exhibits a huge difference
between 2.15~\micron~and $L'$, indicating the specific grain size
responsible for the reddening. To estimate the size of the grain and
the amount of the extinction, we compare the flux density distribution
observed at the cocoon and outside of the cocoon (Fig. \ref{sedge3}).
It is evident the visible light is totally obscured in the cocoon,
while there is no significant color change attributable to the cocoon
in the thermal infrared region, except that the flux density increases
in the cocoon by order of 2. This indicates that the light from the
central region is first scattered by the dust grains in the lobe, and
then subject to extinction in the foreground dust cocoon surrounding
the lobes. The flux density enhancement at thermal wavelengths could
be attributed to the higher particle density closer to the center.

It is obvious that the extinction curve is quite different from the
average interstellar medium. We calculated the extinction efficiency
of amorphous carbon grains of various sizes using the optical
constants for amorphous carbon retrieved from \cite{zub96}.  We found
the reddening property is best reproduced by 0.2~\micron~single-size
spherical amorphous carbon grains, and it has a flat extinction in the
visible wavelength, a large drop between 2 to 3~\micron~region, and an
equally negligible extinction at $L'$ and $M'$. We applied the
reddening by the amorphous carbon to the flux density distribution
observed outside the cocoon with column density of $N_{\rm dust} = 1.5
\times 10^9$~cm$^{-2}$. We found the resulting density distribution
matches that inside the cocoon well (Fig. \ref{ac1}).  The extinction
applied in the cocoon corresponds to $A_V$ = 6.3~mag in visible and
$A_K$ = 2.7~mag in $K$, where the optical depth was estimated from
$\tau_\lambda = N_{\rm dust} Q_{\rm ext} \pi a^2 $ and $a$ is the
particle radius.



\subsection{Dark Lane at the Center of the Nebula}

The smaller extinction at $L'$ allows a view close of the central star
through the heavily obscuring dust cocoon. At this wavelength a dark
lane in the center of the nebula is evident (Fig. \ref{tor1}). This
dark lane, which is seen at $L'$ and $M'$, has a different orientation
from the dust cocoon observed by \citet{sah98b}, and it appears to be
perpendicular to the major axis of the centimeter continuum emission
region observed by \citet{jur00}.

The location of the central star proposed by the polarization map,
searchlights, and the multiple outflow of CO molecules falls close to
the dark lane, but there is no pointlike source at $L'$ or $M'$
corresponding to the central star. We infer the dark lane is a dust
disk lying almost edge-on obscuring the central star. The similar
appearance at $L'$ and $M'$ bands indicates it is absorbing light
evenly at 3--5~\micron. Dust grains in the disk should be cold enough
compared to the surrounding nebulosity to produce no detectable
emission at $L'$ or $M'$. Hence the size of the grain should be quite
large ($a >$ 1~\micron).

Although the nature of the disk is uncertain, there might be a
physical association of the disk orientation to the morphological
features observed in the radio wavelength oriented to PA $\approx$
55$^\circ$ (Fig. \ref{sketch1}). The disk would collimate all these
high-velocity compact components in a direction offset from the
optical polar axis. The high velocity compact components should be the
youngest structures in the nebula. We speculate that the central star
is now experiencing a change in the symmetry axis. Consequently, the
observed position angle of each component near the center may changes
depending on the time it formed. It is tempting to note the series of
blobs in the northern lobe seems to curve from the northwest to the
northeast from the center (Fig. \ref{a2hkl1}). The sequence might be a
tracer of explosive ejection from the central star in the order of
formation. We might be witnessing the central star gradually changing
the ejection axis from PA = 15$^\circ$ to PA $\approx$ 55$^\circ$.

\subsection{The Nature of Peak A}

An enigmatic object near the center of the nebula is peak A, a
pointlike source at the northeast of the central star
(Fig. \ref{tor1}). Based on the distorted polarization pattern near
the peak, \cite{wei00} argued the peak A is a self-luminous infrared
source that is a binary companion to the invisible central star. At
first glance the extremely red color of peak A (Fig. \ref{a2hkl1})
appears to be a clear signature of the thermal emission source, yet we
find evidence to the contrary. We will discuss several observational
features to suggest that the peak A is not self-luminous, but a dense
dusty blob reflecting the infrared light from the central star.

First, although peak A is indeed the brightest spot, the overall look
of the SED does not deviate from the surrounding reflection nebula. To
see this, we compared the SED with those of adjacent regions
(Fig. \ref{peak2}). The infrared (1.65--4.70~\micron) SED of peak A
and those extracted from other locations are almost identical. If peak
A has an illuminating source inside seen through the enshrouding dust
in the thermal infrared, it would be hard to account for the identical
SED in the infrared as far as 2\farcs 3 ($\sim$2800~AU) away from peak
A. The direct light should be much redder than the scattered one. The
only plausible interpretation of the identical SEDs is the identical
radiation mechanism, namely, the relevant region is illuminated by a
single central infrared source, and the reflected light becomes redder
in the foreground dust cocoon. There is obviously no sign of a
pointlike source in the southern lobe where a similar SED is
observed, and thus we conclude peak A is not a self-luminous point
source either, but a dense dusty blob reflecting the illumination from
the central infrared source.

Second, if peak A were indeed self-luminous, what kind of source could
it be? Peak A is located $\sim$600~AU away from the central star in
the projected distance, so it cannot be thermal emission of dust
heated up by the central star ($L_\ast$ = 1.8 $\times$ 10$^4$
$L_\sun$) in equilibrium (thermal equilibrium temperature is
$<$50~K). We may eliminate shock heating, because there is no shocked
feature at the peak A in H$_2$ 2.122~\micron~ narrowband imaging by
\cite{sah98a}. Hence the remaining possibility is an enshrouded
star. The apparent earlier phase of evolution relative to the central
star (no nebulosity association, no hydrogen emission lines) suggests
the possible stellar source should be a G- to M-type main-sequence or
AGB star. However, to date we have no observational support for the
presence of such a late type star there. Spatially resolved
spectroscopy in the near-infrared (1.1--2.4~\micron) made by
\cite{hor94} covers the peak A location, but there is no significant
CO absorption band-head at 2.3~\micron~in their spectra.

Third, our spectroscopy at peak A shows prominent emission features of
hydrocarbon dust. Optically thick thermal emission cores are typically
featureless, or at most exhibit molecular absorption lines in the
3.0~\micron~region due to C$_{\rm 2}$H$_{\rm 2}$ or HCN
\citep{got97,rid78} that are not apparent in our spectrum. As no AGB
stars have been reported to have IEF in the 3~\micron~region, the
hydrocarbon features are a consequence of scattered light originating
from the central star.

Fourth, the observed hydrocarbon features are all uniform around peak
A. The spectra presented in Figure \ref{spec1} are normalized to the
power-law continuum to see if there is any spatial variation in the
hydrocarbon feature at 3.3 and 3.4~\micron, and this is shown in
Figure \ref{spec1_nrm}.  For comparison with the spectral feature
extracted at each location along the slit, a template spectra is
created by combining all the data available.  We found all the
normalized spectra are very similar, and there is no significant
deviation from the template at any location within the observed area.
This is again consistent with the view that the IEF emission source is
remotely illuminating the dusty blobs, and we are observing the
scattered light in the 3~\micron~region.

When scattered light is dominant at 3~\micron, it is surely so at the
shorter wavelengths. However, the polarization map at 2.2~\micron~
presented by \cite{wei00} shows lower polarization close to peak A. We
suspect that multiple scattering in a dense dust blob can reproduce
the same polarization pattern, namely, less polarization at the
brighter location. Polarization imaging in the thermal infrared band,
where the multiple scattering effects are smaller, would be more
conclusive. We conclude that peak A is probably not a self-luminous
source, but is instead one of the dust blobs reflecting infrared
illumination from the central source.

\section{Summary}
We presented high-resolution ($\sim$0\farcs 3) $L'$ and $M'$ images of
AFGL~2688. Imaging in the thermal infrared wavelength at high angular
resolution proved to be a powerful technique to penetrate deep into
the central region of a bipolar proto-planetary nebula with heavy
extinction and that is invisible at shorter wavelengths. We conclude
the following:

1. A dust cocoon described by \cite{sah98b} appears as a very red
region in the south lobe with a dramatic color change between 2 to
3~\micron. The amount of extinction in the dust cocoon was estimated
to be $A_{V}$ = 6.3~mag, assuming 0.2~\micron~amorphous carbon grains
are responsible for the extinction.

2. We found a dark lane at the core of the nebula at PA = 140$^\circ$.
The normal position angle is more consistent with a high-velocity
outflows rather than an optical reflection nebula. The central star is
invisible at both $L'$ or $M'$.

3. Peak A, postulated to be a self-luminous object, has an almost
identical SED to the nearby heavily obscured regions, and we conclude
that it is not self-luminous.

4. From the 3~\micron~spectra we find that the IEFs at the bright
blobs look uniform, and exhibit no significant spatial variation in
the feature-to-continuum ratio. This supports the idea that the bright
blobs are reflecting light from the invisible central star.

\acknowledgments 

We are grateful to M. Weber and R. Potter for their assistance when
these observations were made. We appreciate many useful discussions
with H. Suto. We thank the staff and crew of the Subaru Telescope for
their valuable assistance in obtaining these data. M. Goto is
supported by Japan Society for the Promotion of Science fellowship.

\clearpage
\figurenum{1} \figcaption{Image of AFGL~2688 at $L'$
(3.8~\micron). North is up and East is the left. Left: shown in linear
scale. The peak surface brightness and the total magnitude of the
bipolar nebula are 6.2~mag~arcsec$^{-2}$, and 5.5~mag,
respectively. The spatial resolution is about 0\farcs 3. The image is
convolved with a 2D Gaussian filter of 1.4~pixel FWHM to show the
faint structure better. Right: the same, but shown in logarithmic
scale. \label{imglp1}}

\figurenum{2} \figcaption{The same as Fig. \ref{imglp1}, but at $M'$
(4.7~\micron). The peak surface brightness and total magnitude are
4.4~mag~arcsec$^{-2}$ and 3.5~mag, respectively. The seeing was about
the same as at $L'$, but the actual spatial resolution is somewhat
inferior because of the slightly out-of-focus 22~mas~pixel$^{-1}$
camera at $M'$. The images are convolved with a 2D Gaussian filter of
1.4~pixel FWHM. \label{imgmp1}}

\figurenum{3} \figcaption{True-color composite image of
AFGL~2688. Blue, green, and red represent $H$ (1.65~\micron) and H$_2$
continuum (2.15~\micron) obtained with {\it HST/NICMOS}, and $L'$ with
the IRCS. Left: The image is normalized to the maximum intensity of
the each band before combining into a color image. Right: The same,
but with a different stretch to emphasize the color change
1\arcsec~south from the tip of the southern lobe. The encircled
regions are the locations of aperture photometry in
Fig. \ref{sedge3}. \label{a2hkl1}}

\figurenum{4} \figcaption{$L'$ contour plot of AFGL~2688 with an overlay
showing the slit position for the grism spectroscopy.\label{slit2} }

\figurenum{5} \figcaption{Spatial variation in the 3~\micron~spectra
of AFGL~2688 at the three locations illustrated in
Fig. \ref{slit2}. The spectra are extracted at every 0\farcs 233
(=~4~pixel) along the slit. The off-center spectra are offset by 2
$\times$ 10$^{-14}$~[W~m$^{-2}$~\micron$^{-1}$] for clarity. The
continua are defined by fitting both sides of the emission features
(2.95--3.15~\micron~and 3.60--3.80~\micron) with a power-law
function. \label{spec1}}

\figurenum{6} \figcaption{A schematic view summarizing the major
morphological features of AFGL~2688. The structural components
described here can be divided into two distinct groups, one aligned to
PA = 15$^\circ$ and the other to 53$^\circ$--60$^\circ$. The
concentric arcs, the searchlights, the delineated bipolar lobes, and
the dust cocoon all detected at visible to near-infrared wavelengths
\citep{sah98a,sah98b,lat93} are aligned to PA = 15$^\circ$. The
high-velocity (26.5~km~s$^{-1}$ at projected velocity) component of
the $^{13}$CO outflow \citep{yam95}, the compact expanding CO shell
($>$10~km~s$^{-1}$) \citep{cox00}, and the centimeter wavelength
continuum emission \citep{jur00} are aligned to PA =
53$^\circ$--60$^\circ$. The low-velocity wind (10--20~km~s$^{-1}$) is
thought to be a remnant of spherical mass loss in the preceding AGB
phase. The moderate-velocity wind ($\sim$ 40~km~s$^{-1}$) might be
associated to the near-infrared bipolar nebula invading the
low-velocity wind \citep{you92}. Shocked H$_2$ molecule blobs are
distributed orthogonally about the center to cap the optical bipolar
lobes at the tips. The CO multiple outflows apparently originating at
the central source are aligned with the H$_2$ blobs but are not shown
here \citep{cox00}. \label{sketch1}}

\figurenum{7} \figcaption{Left: Crosscuts of the surface brightness of
AFGL~2688 at the south edge at $H$ (1.65~\micron), and the H$_2$
continuum (2.15~\micron), $L'$, and $M'$. The abscissa is an offset
from the center along the optical bipolar axis (PA = 15$^\circ$). The
location of the edge where the surface brightness in the 2.15~\micron~
sharply drops is marked with solid lines ($\sim$1\arcsec~from the
center). Right: A crosscut of the color variation ($H-$2.15~\micron,
2.15~\micron$-L'$, and $L'-M'$) in magnitude unit at the same location
as the left panel. \label{sedge1}}

\figurenum{8} \figcaption{The aperture photometry inside and outside
the dust cocoon. The aperture locations are marked in
Fig. \ref{a2hkl1}.  The surface brightness outside the cocoon is
reddened again by a 0.2~\micron~amorphous carbon grain to estimate the
amount of extinction. The result is well matched to the surface
brightness inside the cocoon after applying the extinction by $N_{\rm
dust}$ = 1.5 $\times$ 10$^9$~cm$^{-2}$.\label{sedge3}}

\figurenum{9} \figcaption{An extinction efficiency of an amorphous
carbon grain. The extinction efficiency is calculated with Mie theory
for a single-size ($a$ = 0.1~\micron~to $a $ = 2.0~\micron) spherical
amorphous carbon grain at the specific wavelengths of the {\it
HST/WFPC2}, {\it NICMOS} and IRCS filter systems. The optical
constants are taken from BE amorphous carbon of \cite{zub96}.
Monochromatic extinction in the visible, a huge extinction drop
between 2 to 3~\micron, and less extinction at longer than $L'$ are
best reproduced by 0.2~\micron~size grains. \label{ac1}}

\figurenum{10} \figcaption{A blow up of the central 4\arcsec $\times$
4\arcsec~region of AFGL~2688 at $L'$. The orientation of the dark lane
is outlined by dashed lines to contrast with the axis of optical
bipolar nebula. The dotted lines trace the local minimum of the
image. The locations of peak A and the central star (source B) are
marked with circles. The position of the central star is from the
polarization mapping of \citet{wei00} and multiple CO outflow
observations of \citet{cox00}. The formal uncertainty in the position
of source B by \citet{wei00} is $\lesssim$ 0\farcs 03, and the
uncertainty in the radio observation by \citet{cox00} is
$\lesssim$1\arcsec. \label{tor1}}

\figurenum{11} \figcaption{The near-infrared SED of peak A and the
adjacent regions. A series of aperture photometry was made at the
both sides of peak A from 0\farcs 5 southeast to 2\farcs 3 to
northeast with 10 pixels (0\farcs 583) aperture and
separation. \label{peak2}}

\figurenum{12}\figcaption{The 3~\micron~spectra of AFGL~2688
normalized to the power-law continuum defined in Fig. \ref{spec1}. The
overlay is a template spectrum to show how similar all of the
spectra are. \label{spec1_nrm}}

\newpage



\begin{figure}
\figurenum{4}
\plotone{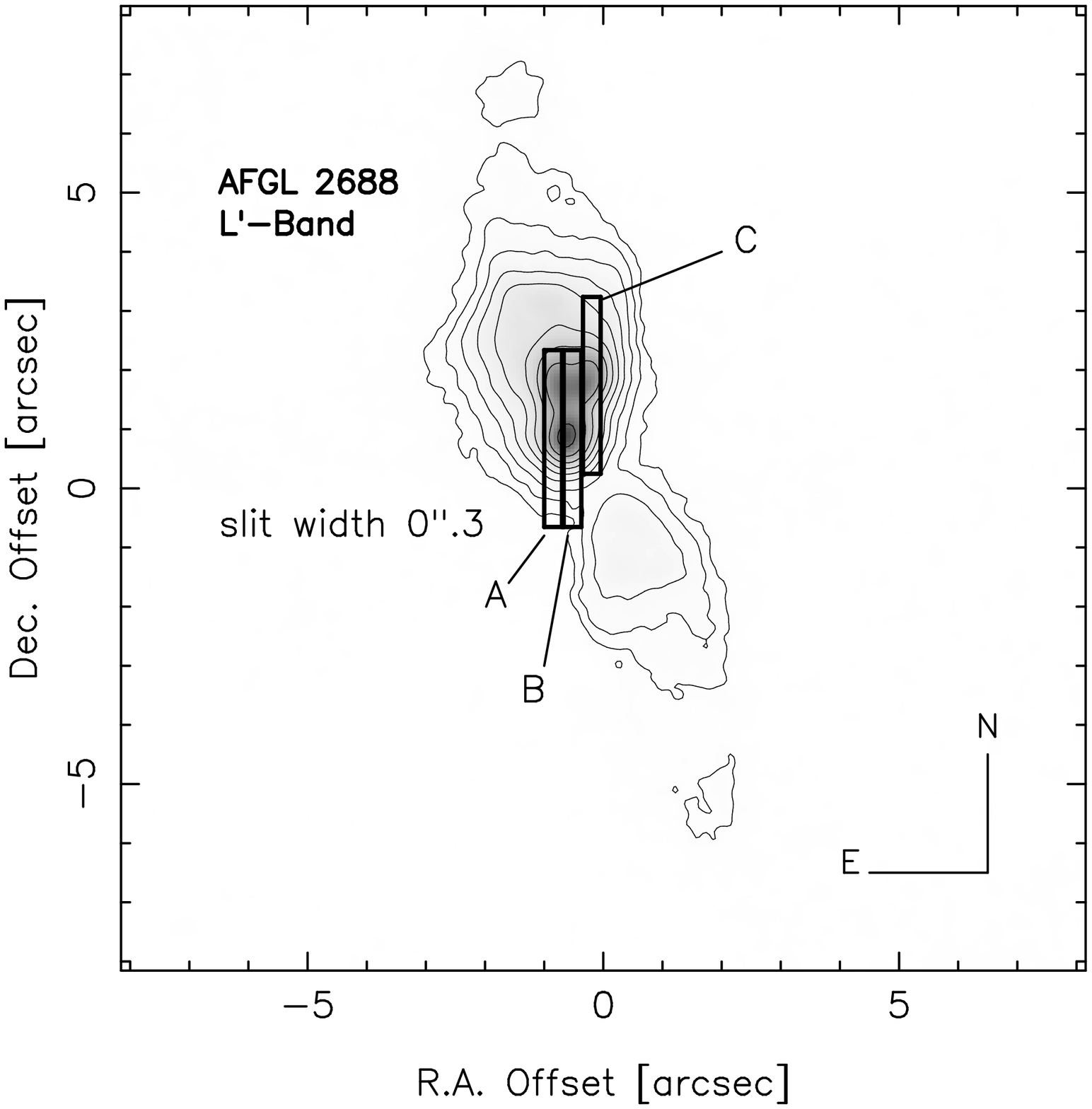}
\caption{}
\end{figure}

\begin{figure}
\figurenum{5}
\plotone{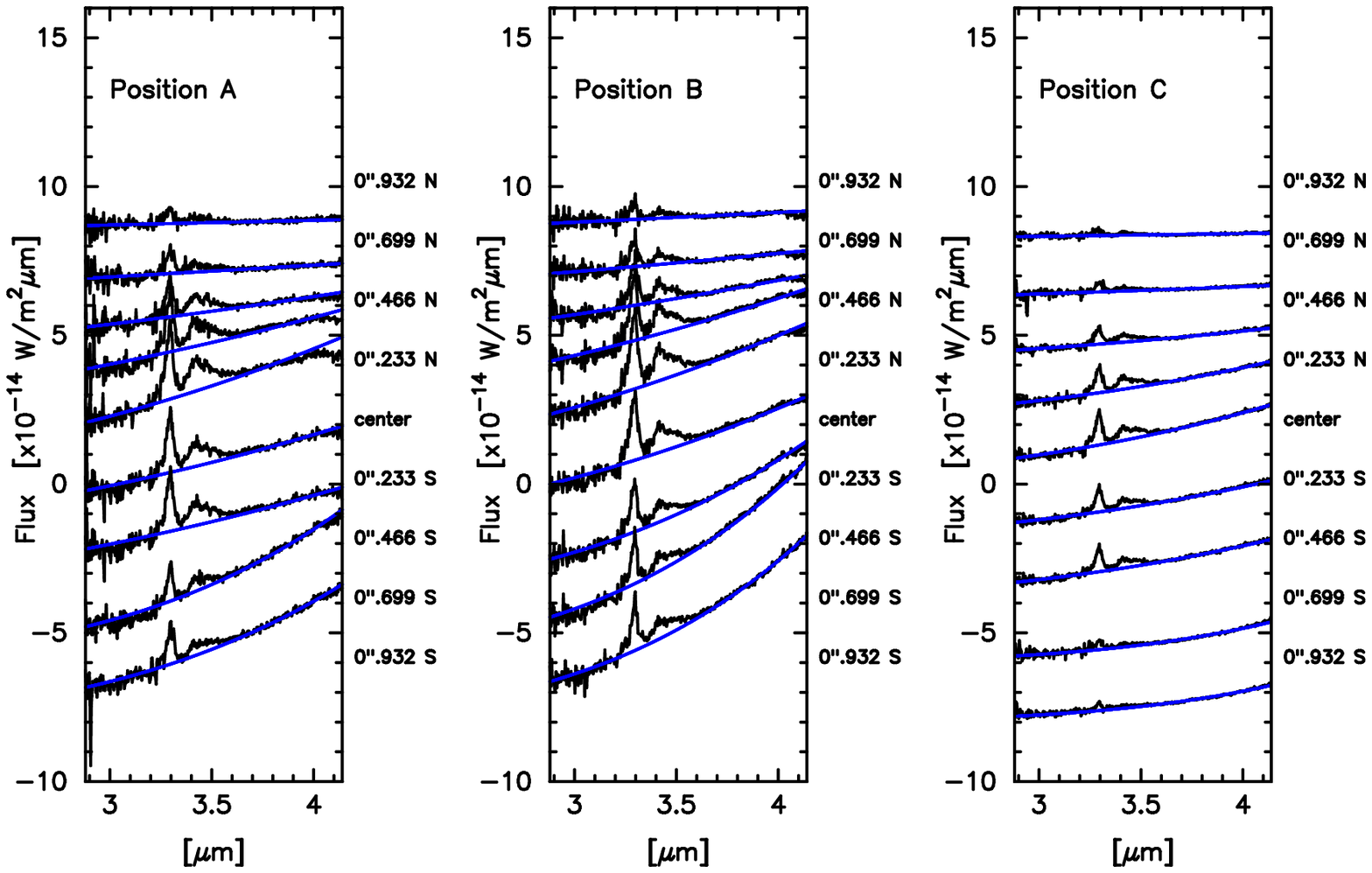}
\caption{}
\end{figure}


\begin{figure}
\figurenum{7}
\plotone{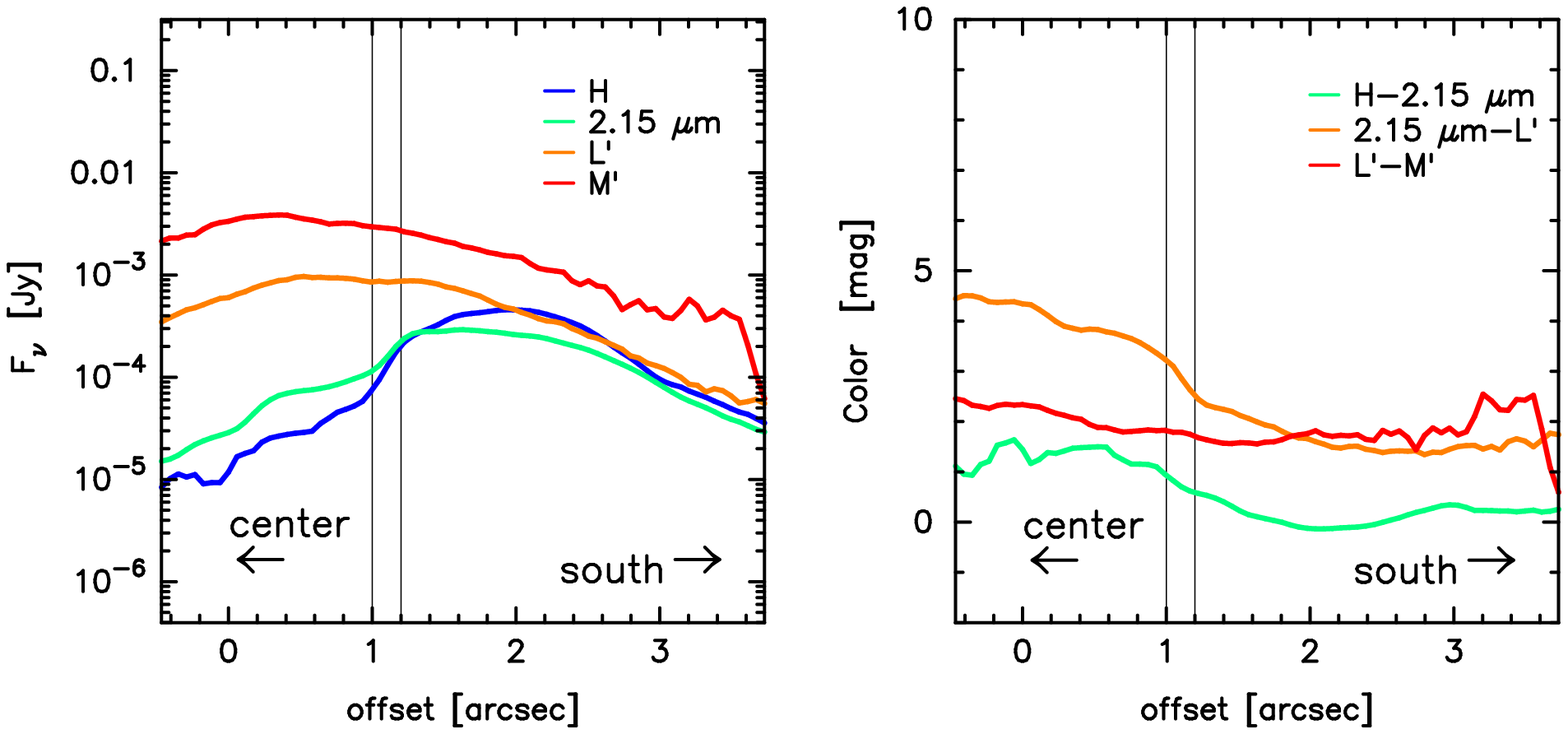}
\caption{}
\end{figure}

\begin{figure}
\figurenum{8}
\plotone{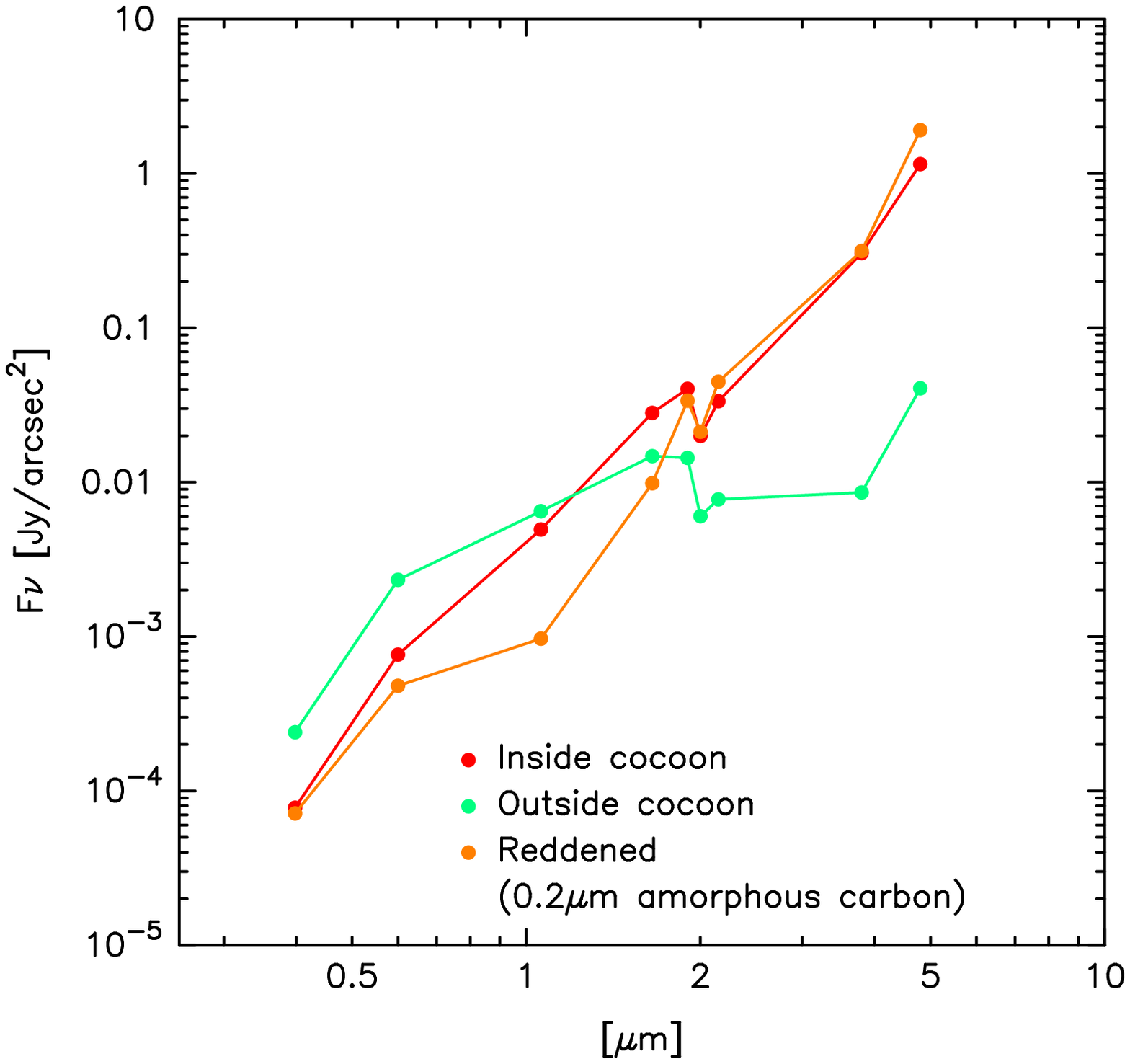}
\caption{}
\end{figure}

\begin{figure}
\figurenum{9}
\plotone{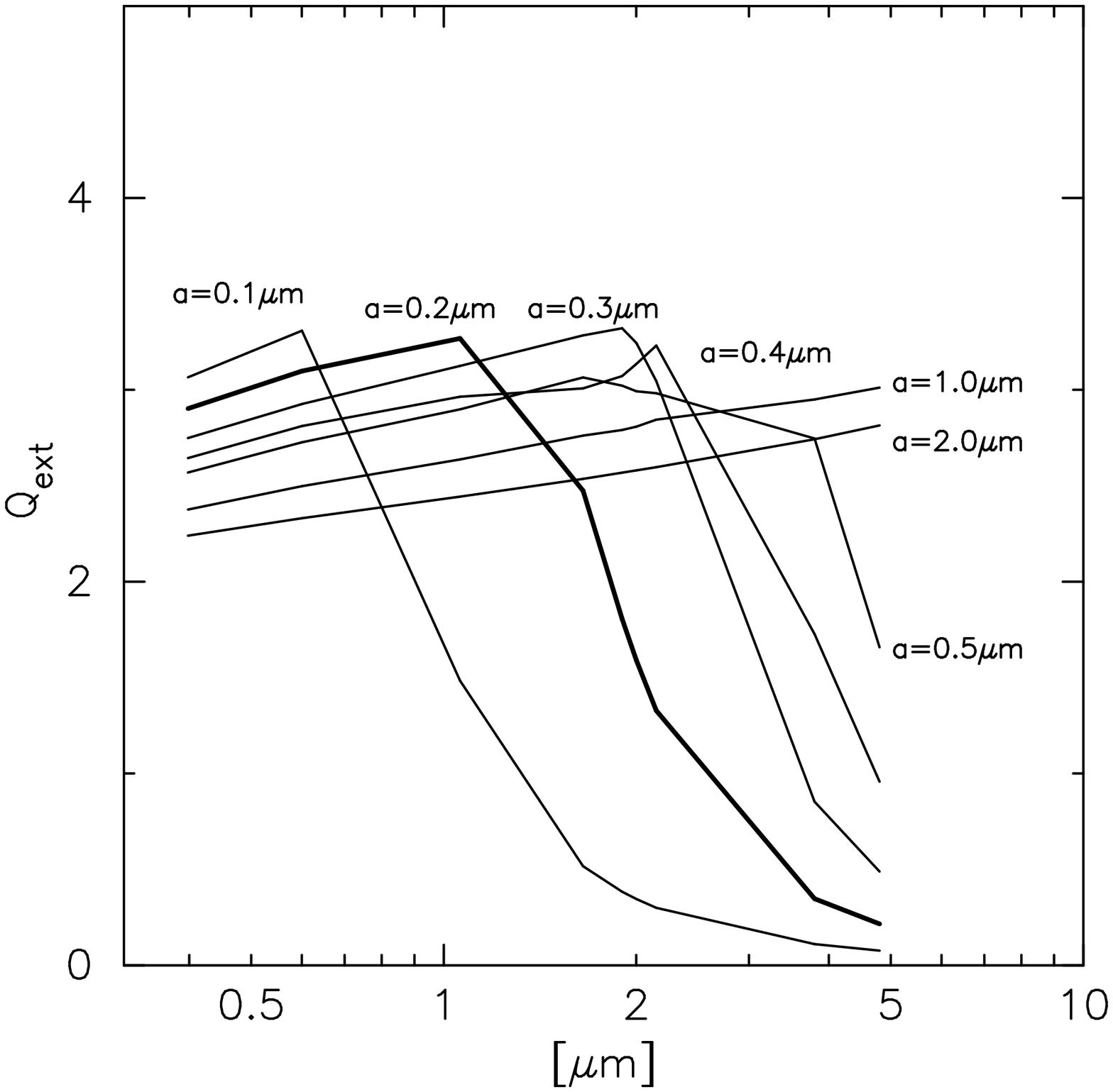}
\caption{}
\end{figure}


\begin{figure}
\figurenum{11}
\plotone{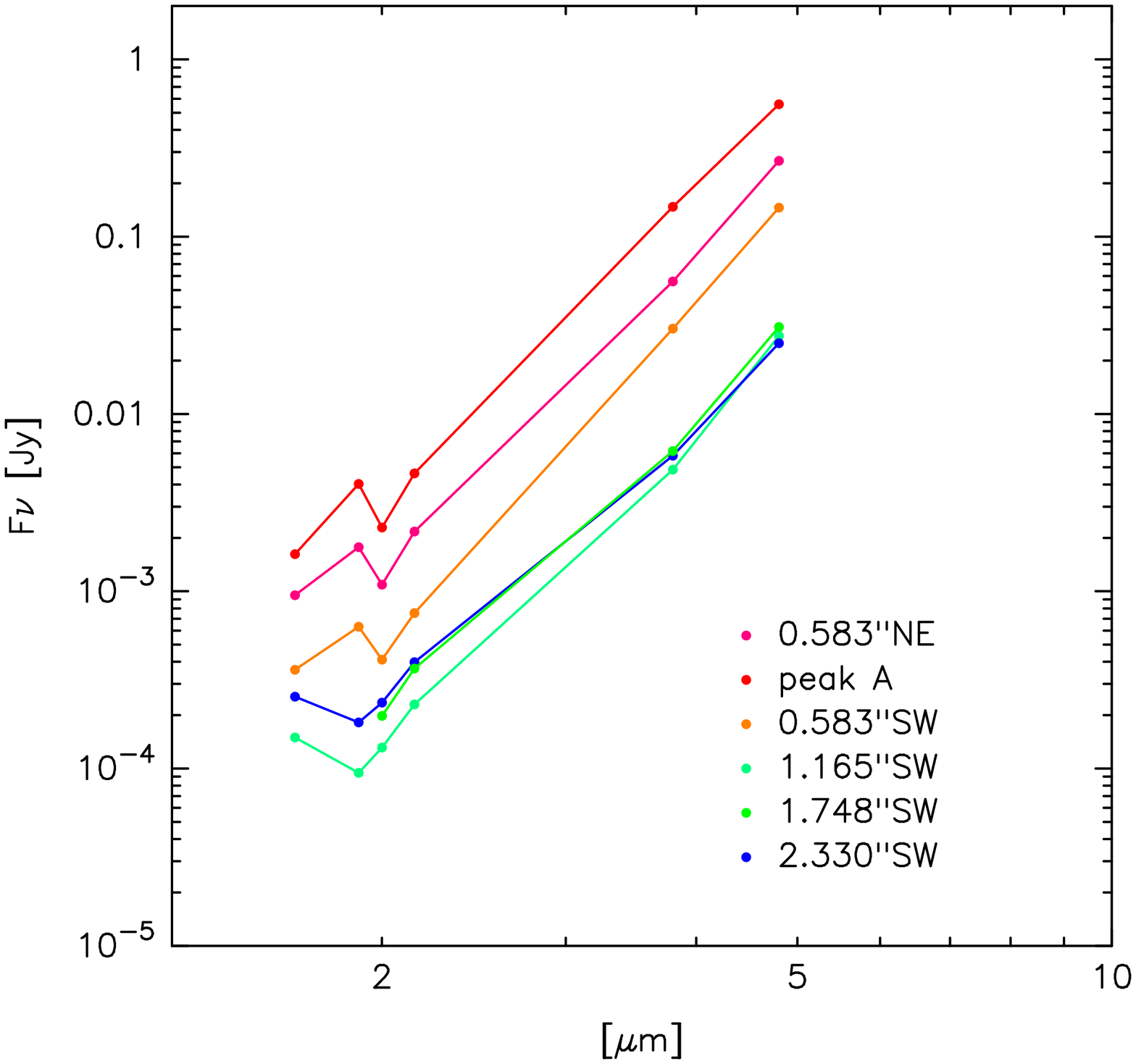}
\caption{}
\end{figure}

\begin{figure}
\figurenum{12}
\plotone{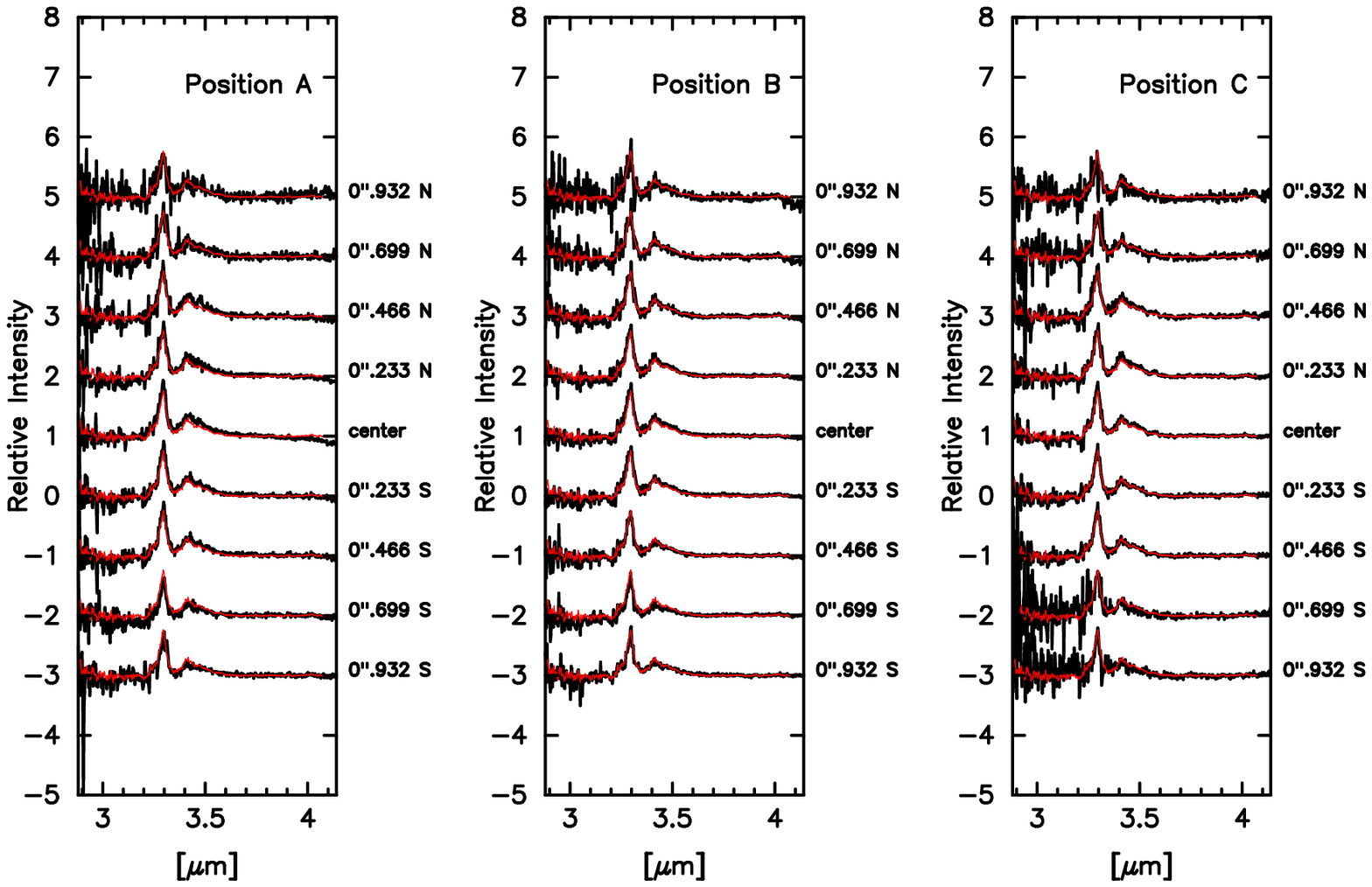}
\caption{}
\end{figure}

\end{document}